\documentclass[AMA,STIX2COL,Linenumberson]{MRM}
\articletype{Research article}%

\received{01 Dec 2023}
\revised{31 Mar 2024}
\accepted{01 Apr 2024}
\usepackage{xspace}
\usepackage{setspace}
\usepackage{amssymb}
\usepackage{amsmath}
\usepackage{multirow}
\usepackage{comment}

\newcommand{\textr}[1]{#1} 

\newcommand{\core}{CORe\xspace}

\renewcommand{\eqref}[1]{Equation~(\ref{eq:#1})}

\newcommand{\tabref}[1]{Table~\ref{tab:#1}\!\!\!}
\newcommand{\figref}[1]{Figure~\ref{fig:#1}\!\!\!} 

 \newcommand{\algref}[1]{Algorithm~\ref{alg:#1}}
 \newcommand{\linref}[1]{Line~\ref{lin:#1}}

\newcolumntype{L}[1]{>{\raggedright\let\newline\\\arraybackslash\hspace{0pt}}m{#1}}
\newcolumntype{C}[1]{>{\centering\let\newline\\\arraybackslash\hspace{0pt}}m{#1}}
\newcolumntype{R}[1]{>{\raggedleft\let\newline\\\arraybackslash\hspace{0pt}}m{#1}}

\DeclareMathOperator*{\argmin}{arg\,min}

\DeclareMathOperator{\real}{Re}

\renewcommand{\Tilde}{\widetilde}
\renewcommand{\Hat}{\widehat}

\renewcommand{\vec}[1]{\ensuremath{\boldsymbol{#1}}}

\newcommand{\hvec}[1]{\ensuremath{\Hat{\boldsymbol{#1}}}}

\newcommand{\tran}{^{\top}}
\newcommand{\herm}{^\textsf{H}}
\newcommand{\Real}{{\mathbb{R}}}
\newcommand{\Complex}{{\mathbb{C}}}

\begin{document}

\title{Motion-robust free-running \textr{volumetric} cardiovascular MRI}

\author[1,2]{Syed M. Arshad}{}
\author[2,3]{Lee C. Potter}{}
\author[1,2]{Chong Chen}{}
\author[3]{Yingmin Liu}{}
\author[3]{Preethi Chandrasekaran}{}
\author[4]{Christopher Crabtree}{}
\author[5]{Matthew S. Tong}{}
\author[3,5]{Orlando P. Simonetti}{}
\author[5]{\\Yuchi Han}{}
\author[1,2,3]{Rizwan Ahmad}{}

\authormark{ARSHAD \textsc{et al.}}

\address[1]{\orgdiv{Biomedical Engineering}, \orgname{The Ohio State University}, \orgaddress{\state{Ohio}, \country{USA}}}

\address[2]{\orgdiv{Electrical \& Computer Engineering}, \orgname{The Ohio State University}, \orgaddress{\state{Ohio}, \country{USA}}}

\address[3]{\orgdiv{Davis Heart and Lung Research Institute}, \orgname{The Ohio State University Wexner Medical Center}, \orgaddress{\state{Ohio}, \country{USA}}}

\address[4]{\orgdiv{Human Sciences}, \orgname{The Ohio State University}, \orgaddress{\state{Ohio}, \country{USA}}}

\address[5]{\orgdiv{Internal Medicine}, \orgname{The Ohio State University Wexner Medical Center}, \orgaddress{\state{Ohio}, \country{USA}}}

\corres{Rizwan Ahmad, \email{ahmad.46@osu.edu}}

\presentaddress{Biomedical Research Tower, 460 W 12th Ave, Room 318, Columbus OH 43210, USA.}

\finfo{This work was partially supported by \fundingAgency{National Institute of Health} grants \fundingNumber{R01HL135489} and \fundingNumber{R01HL151697}}

\abstract[Abstract]{\section{Purpose} \textr{To present and assess an outlier mitigation method that makes free-running volumetric cardiovascular MRI (CMR) more robust to motion.}
\section{Methods} The proposed method, called Compressive recovery with Outlier Rejection (CORe), models outliers in the measured data as an additive auxiliary variable. We enforce MR physics-guided group sparsity on the auxiliary variable, and jointly estimate it along with the image using an iterative algorithm. For evaluation, CORe is first compared to traditional compressed sensing (CS), robust regression (RR), and an existing outlier rejection method using two simulation studies. Then, CORe is compared to CS using seven 3D cine, twelve rest 4D flow, and eight stress 4D flow imaging datasets. 
\section{Results} Our simulation studies show that CORe outperforms CS, RR, and the existing outlier rejection method in terms of normalized mean square error (NMSE) and structural similarity index (SSIM) across 55 different realizations. The expert reader evaluation of 3D cine images demonstrates that CORe is more effective in suppressing artifacts while maintaining or improving image sharpness. Finally, 4D flow images show that CORe yields more reliable and consistent flow measurements, especially in the presence of involuntary subject motion or exercise stress.
\section{Conclusion} \textr{An outlier rejection method is presented and tested using simulated and measured data. This method can help suppress motion artifacts in a wide range of free-running CMR applications.}}

\jnlcitation{
\ctitle{Arshad SM, Potter LC, Chen C, et al. Motion-robust free-running volumetric cardiovascular MRI. Magn Reson Med. 2024; 92(3): 1248-1262. doi: 10.1002/mrm.30123}
}

\maketitle
\section{Introduction}\label{sec:intro}
Magnetic resonance imaging (MRI) is a versatile imaging modality that provides high soft-tissue contrast without ionizing radiation. Cardiovascular magnetic resonance imaging (CMR) provides comprehensive assessment of the cardiovascular structure, function, and morphology. CMR-based assessment is unmatched in assessing cardiac function, detecting myocardial fibrosis, and providing comprehensive tissue characterization.\cite{leiner2020scmr}

In addition to occasional bulk motion, CMR scans must account for two physiological motions: cardiac and respiratory. Most clinical CMR scans are based on breath-held segmented acquisition, where the data collection is synchronized with cardiac activity recorded using an electrocardiogram (ECG).\cite{brookeman1986bhmri,edelman1991bhmri} This approach is not feasible for subjects who cannot hold their breath or have arrhythmias. Therefore, 2D free-breathing real-time imaging has gained popularity as it does not require breath-holding or a regular cardiac rhythm.\cite{stuber1997realtime,kerr1998realtime} However, due to a 6 to 10 mm slice thickness and difficulty in temporally aligning different slices, 2D imaging does not yield a true 3D visualization of heart anatomy. Free-breathing volumetric CMR circumvents the limitations of 2D imaging, but the respiratory motion of the heart in free-breathing acquisition remains a major challenge.\cite{ehman1984freebreathing} Free-breathing volumetric CMR performed under ECG guidance and prospective respiratory gating using navigator echoes has been proposed and validated for many CMR applications. However, depending on the breathing pattern and the extent of arrhythmia, this approach may lead to unpredictably long acquisition times. \textr{Also, navigator echoes disrupt the steady-state of magnetization and thus are not compatible with several common CMR pulse sequences. 
More recently, there has been renewed interest in free-running volumetric imaging (FRV), where data are collected continuously for several minutes without guidance from navigator echoes, respiratory bellows, \cite{uribe2007whole} or even ECG.\cite{di2019automated} FRV provides an easier clinical setup with minimal planning and the added flexibility of determining the number of cardiac and respiratory bins at the time of reconstruction.}

A key ingredient in most recent FRV methods is self-gating, \cite{larson2005preliminary} where a fixed segment of k-space--typically a readout in the inferior-superior direction--is traversed periodically during the acquisition. The dynamic changes in self-gating data segments are attributed to physiological motions. A common approach for extracting respiratory and cardiac signals relies on 
performing blind source separation by a combination of band-pass filtering, principal component analysis (PCA), and independent component analysis (ICA).\cite{di2019automated} These motion signals are then employed to sort k-space data into multiple respiratory and cardiac motion bins. \textr{The data binning is then followed by image reconstruction in the spatial-cardiac-respiratory domain, leading to ``motion-resolved'' imaging. \cite{feng2016xdgrasp,feng20185dheart} 
An alternative to motion-resolved imaging is to integrate a respiratory motion model into the reconstruction framework to map images at different respiratory states to a target (e.g., expiratory) state.} \cite{bhat2011affine,dwight2014motioncorr} This approach does not resolve the respiratory dimension but utilizes all the data to improve the image quality of the target respiratory state. However, the efficacy of all these FRV imaging methods depends on the quality of extracted motion signals.

Accurately extracting cardiac and respiratory signals from the self-gating data is a challenging problem, especially when heart-rate variability, arrhythmia, or inconsistent breathing patterns are present. Although Pilot Tone (PT) provides an alternative to self-gating,\cite{ludwig2021pilot} the extraction of physiological motion from PT faces similar challenges.\cite{chen2023PT} Any inaccuracy in the extraction of physiological motion signals would result in incorrect assignment of k-space lines to cardiac and respiratory bins, resulting in image artifacts. In the case of in-magnet exercise stress CMR, the problem of reliably extracting respiratory and cardiac signals becomes even more challenging due to the movement of the torso.\cite{gerche2013exercise}

In this work, we introduce a new method called Compressive recovery with Outlier Rejection (\core). In contrast to standard compressed sensing (CS),\cite{lustig2008} \core provides more robust reconstruction by suppressing the outliers, 
which are invariably present in FRV acquisitions due to imperfect binning. \core explicitly models the outliers using an additive auxiliary variable, which is then jointly estimated with the image. Unlike previous work in image inpainting where such an auxiliary variable is assumed to be pixel-wise sparse,\cite{dong2012wavelet} our specific contribution includes leveraging the structure in the MRI data to impose sparsity at a group (readout) level. Additionally, we implement \core using the alternating direction method of multipliers (ADMM) optimization algorithm and apply it to large-scale imaging problems. We have tested our method using two simulation studies as well as data from free-running 3D cine, 4D flow, and exercise stress 4D flow imaging.

\section{Theory}\label{sec:the}
\textr{In MRI, reconstruction involves estimating the underlying image from noisy and potentially undersampled complex-valued k-space measurements.} The measured noisy data are related to the image by
\begin{equation}
\label{eq:model}
\vec{y} = \vec{A}\vec{x} + \vec{\omega},
\end{equation}
where $\vec{x}\in\Complex^N$ is an $N$-pixel image that has been vectorized, $\vec{y}\in\Complex^M$ is the MRI data measured from $C$ receive coils, $\vec{\omega}\in\Complex^M$ is circularly symmetric white Gaussian noise with variance $\sigma^2$, and $\vec{A}\in \Complex^{M\times N}$ is a known sensing matrix that incorporates pixel-wise multiplication with coil sensitivity maps, 2D or 3D discrete Fourier transform, and k-space undersampling. 
The model in \eqref{model} is also applicable to dynamic imaging where $\vec{x}$ and $\vec{y}$ represent vertical concatenations of pixels and k-space data, respectively, from individual frames, and $\vec{A}$ represents block-diagonal embedding of the sensing matrices from individual frames. 

\textr{In CS-based reconstruction,\cite{lustig2007} \eqref{model} is often solved using
\begin{equation}
\label{eq:map1}
\hvec{x}_{\text{CS}} =  \argmin_{\vec{x}}\frac{1}{\sigma^2}\|\vec{Ax}-\vec{y}\|_2^2 + \mathcal{R}(\vec{x}),
\end{equation}}
\textr{where $\mathcal{R}(\vec{x})$ is a sparsity promoting prior. Common choices of $\mathcal{R}(\vec{x})$ in CS-based MRI reconstruction include $\lambda_1\|\vec{\Psi}\vec{x}\|_1$, where $\vec{\Psi}$ is a linear sparsifying transform,\cite{abstractlustig2007} or $\lambda_1\mathcal{T}(\vec{x})$, where $\mathcal{T}(\cdot)$ computes the total variation.\cite{ROF92}} 
From a Bayesian perspective, a regularizer injects prior belief about the underlying image, $\vec{x}$. For example, for $\mathcal{R}(\vec{x})=\lambda_1\|\vec{\Psi}\vec{x}\|_1$, $\hvec{x}_{\text{CS}}$ is a maximum a posteriori (MAP) estimate under the sparsity-promoting prior probability density function of $p(\vec{x}) \propto \exp(-\lambda_1\|\vec{\Psi}\vec{x}\|_1)$. 

Several studies have shown that CS methods enable higher acceleration rates than possible without sparsity-based priors,\cite{lustig2008, jaspan2015} and several such reconstruction methods are available on commercial scanners. More recently, deep learning (DL)-based reconstruction methods have been shown to outperform sparsity-based CS methods.\cite{vishnevskiy2020,ravishankar2019image,bustin2020compressed} However, almost all of these CS and DL methods are reliant on \eqref{model}, which, in the absence of motion, is a valid model for MRI measurements. 

In the presence of uncompensated motion, e.g., due to imperfect retrospective data binning, the model in \eqref{model} is no longer valid. If one were to consider motion-induced outliers to be additive noise with a heavy-tailed Laplacian distribution, then a common remedy is to use robust regression (RR) by replacing the $\ell_2$-norm with $\ell_1$-norm in \eqref{map1},\cite{Nikolova2004} leading to the MAP solution
\begin{equation}
\label{eq:map2}
\hvec{x}_{\text{RR}} =  \argmin_{\vec{x}}\lambda_0\|\vec{Ax}-\vec{y}\|_1 + \mathcal{R}(\vec{x}).
\end{equation}

\textr{A different approach to account for motion induced outliers is to model them as an additive perturbation in k-space using an unknown auxiliary variable, $\vec{v}\in\Complex^{M}$, i.e., }
\begin{equation}
\label{eq:model-v}
\vec{y} = \vec{A}\vec{x} + \vec{v} + \vec{\omega}.
\end{equation}
\textr{The model in \eqref{model-v} implies that the underlying image $\vec{x}$, additive Gaussian noise $\vec{\omega}$, and uncompensated motion $\vec{v}$ all contribute to the k-space measurements. However, during image reconstruction process, $\vec{v}$ is not treated as noise but as an auxiliary unknown variable that exhibits structure from the data acquisition and is jointly estimated with $\vec{x}$.} With both $\vec{x}$ and $\vec{v}$ being unknown, the model in \eqref{model-v} is ill-posed even in the presence of $\mathcal{R}(\vec{x})$. To mitigate this issue, one could enforce sparsity on $\vec{v}$ by assuming that its entries are drawn from i.i.d. Laplacian distribution, i.e., $p(\vec{v}) \propto \exp(-\lambda_2\|\vec{v}\|_1)$. This assumption encourages $\vec{v}$ to assume a sparse support in k-space, with its non-zero entries acting as outliers. The resulting optimization problem, termed as sparse outlier (SO), assumes the form
\begin{equation}
\label{eq:l2l1v}
\hvec{x}_{\text{SO}} =  \argmin_{\vec{x},\vec{v}}\frac{1}{\sigma^2}\|\vec{Ax}-\vec{y}+\vec{v}\|_2^2 + \mathcal{R}(\vec{x}) + \lambda_2\|\vec{v}\|_1.
\end{equation}
In \eqref{l2l1v}, although both $\hvec{x}_\text{SO}$ and $\hvec{v}$ are returned, the latter is ignored for brevity. Under mild assumptions, it is easy to show that \eqref{l2l1v} is a MAP estimate of the model in \eqref{model-v}; see Section (A) in Appendix. 

Although the optimization problem in \eqref{l2l1v} has previously been proposed by Dong et al. for blind image inpainting,\cite{dong2012wavelet} it does not fully utilize the data structure specific to MRI measurements. The MRI data are invariably collected in the form of readouts, where a sequence of k-space samples is collected within a short period of time that is of the order of a millisecond. Given that physiological motions occur at much larger time scales, the duration of a single readout can be considered negligible. Therefore, instead of assuming that the motion impacts individual samples in k-space as in \eqref{l2l1v}, we assume that the motion impacts an entire readout. This modeling choice is not only more realistic but also more robust, as it is easier to detect an entire readout corrupted by motion compared to individual k-space entries.

To treat all $K$ k-space samples in each of the $J$ readouts as a group, we leverage the idea of group sparsity.\cite{Bach2008,huang2010} This can be realized by enforcing a hybrid $\ell_2$-$\ell_1$ penalty on $\vec{v}$, where the $\ell_2$ norm is first computed along the readout dimension followed by the $\ell_1$ norm along all remaining dimensions, leading to the proposed \core formulation, i.e., 
 \begin{equation}
\label{eq:core}
\hvec{x}_{\text{CO}} =  \argmin_{\vec{x},\vec{v}}\frac{1}{\sigma^2}\|\vec{Ax}-\vec{y}+\vec{v}\|_2^2 + \mathcal{R}(\vec{x}) + \lambda_2\|g(\vec{v})\|_1.
\end{equation}
In \eqref{core}, the vector $g(\vec{v})$ denotes $\ell_2$-norms of k-space readouts. With $\vec{v}_j\in \mathbb{C}^{K\times 1}$ representing the $j^{\text{th}}$ readout, $g(\vec{v}) \triangleq [\Tilde{v}_1, \Tilde{v}_2,\dots,\Tilde{v}_J] \tran \in \Real^{J\times 1}$, where $\Tilde{v}_j = \|\vec{v}_j\|_2 \in \Real$ and $\|g(\vec{v})\|_1=\sum_{j=1}^J |\Tilde{v}_j|$. As previously, although both $\hvec{x}_\text{CO}$ and $\hvec{v}$ are returned in \eqref{core}, the latter is ignored for brevity. The optimization in \eqref{core} provides automatic separation of measured data into outlier estimation, $\hvec{v}$, and k-space data, $\vec{y}-\hvec{v}$, that is used for image reconstruction. \textr{Note, \core does not make assumptions about the nature of the physiological motion, e.g., rigid vs. nonrigid, or the binning process; it simply employs an additive auxiliary variable as a means to separate contributions from corrupted readouts that are inconsistent with the rest of the data.}

Traditional CS and RR methods require adjusting one regularization parameter, $\lambda_1$, which controls the relative emphasis on $\mathcal{R}(\vec{x})$ compared to the data fidelity term. In contrast, SO and \core require adjusting another parameter, $\lambda_2$, which controls the extent of outlier rejection. A smaller value of $\lambda_2$ corresponds to more aggressive outlier rejection at the potential cost of discarding valid data.

\section{Methods}\label{sec:mat}
For evaluation of the proposed reconstruction method, we performed two simulation studies with retrospective undersampling and three in vivo studies with prospective undersampling. The first simulation study was performed on a static 2D phantom. For the second simulation, we used a digital dynamic phantom. In both simulation studies, we compared the four reconstruction methods: CS (\eqref{map1}), RR (\eqref{map2}), SO (\eqref{l2l1v}), and \core (\eqref{core}) as discussed in Section \ref{sec:the}. After establishing the advantage of \core over CS, RR, and SO in the simulation studies, we compared \core with the conventional CS reconstruction framework (\eqref{map1}) in volumetric CMR reconstruction. In all studies, we chose $\mathcal{R}(\vec{x})=\lambda_1 \|\vec{\Psi}\vec{x}\|_1$, with $\vec{\Psi}$ representing the undecimated wavelet transform in the spatio-temporal domain.\cite{KAYVANRAD2014udwt} In vivo studies were performed on consented healthy subjects and patients, approved by the institutional review board (IRB).

\subsection{Study I--Static phantom}
\begin{figure*}[ht!]
    \begin{center}
    \includegraphics[width = 0.95\textwidth]{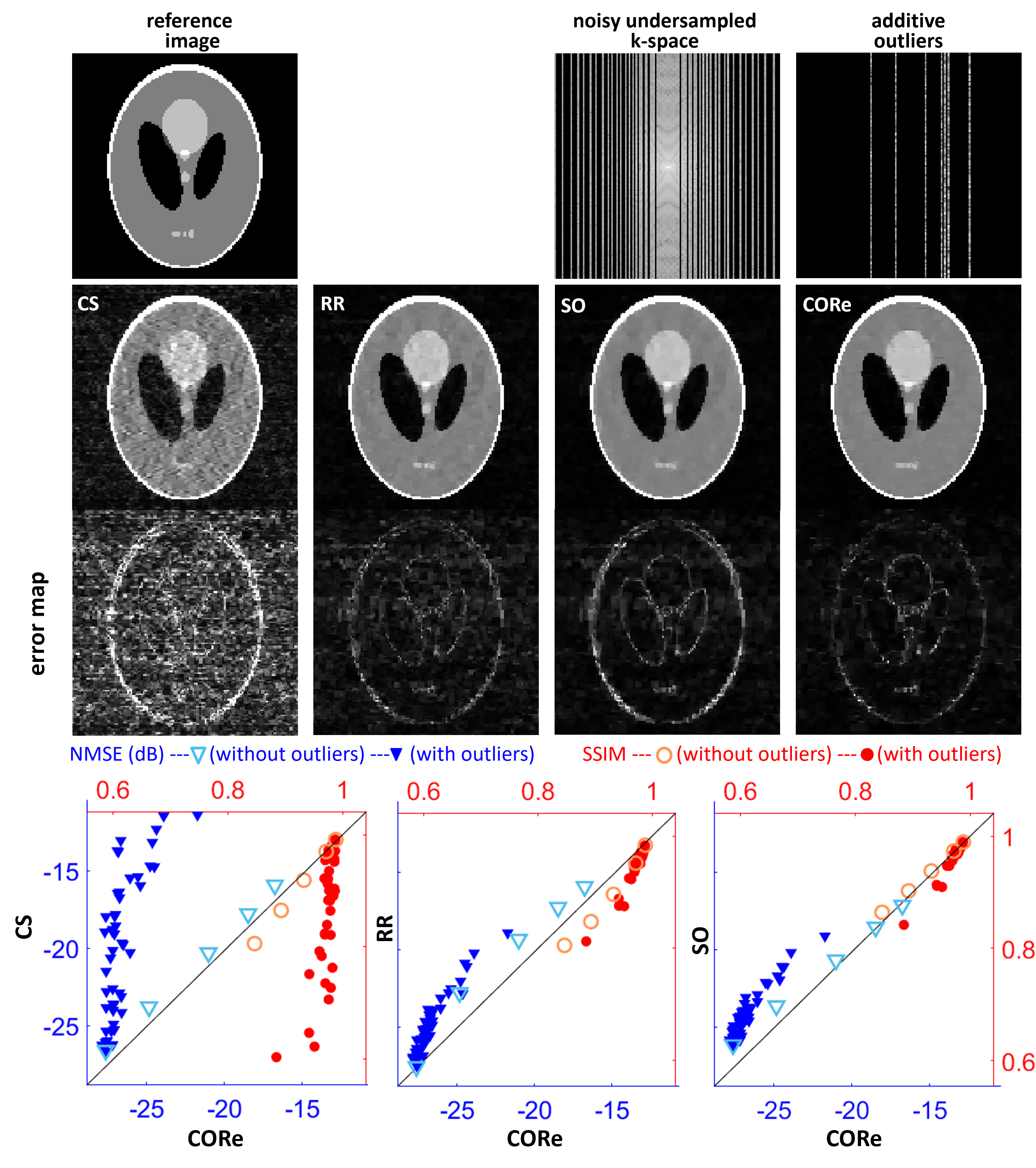}
    \end{center}
    \caption{\textr{Summary of the results from Study I. The top row (from left to right) displays the reference image, noisy undersampled k-space measurements, and additive outliers for a representative realization. The second row shows reconstructed images using CS, RR, SO, and \core. The third row shows $3\times$ amplified error maps of the corresponding reconstructed images. The bottom row displays scatter plots for 55 realizations, comparing \core with CS, RR, and SO in terms of NMSE and SSIM, both with and without outliers.}}
    \label{fig:Figure 1}
\end{figure*}

In this study, we compared CS, RR, SO, and \core for the reconstruction of a $128 \times 128$ Shepp-Logan phantom. \textr{To simulate single-coil k-space data, a 2D discrete Fourier transform of the digital phantom was performed, followed by Cartesian undersampling, using the golden ratio offset (GRO) sampling pattern.\cite{joshi2022GRO}} The net acceleration rate was fixed at $2.2$. To simulate noisy measurements, circularly symmetric white Gaussian noise with a fixed variance $\sigma^2$ was added to the undersampled k-space data, as shown in \figref{Figure 1}. \textr{To introduce outliers, a certain fraction of readouts was contaminated with additional noise. The severity of these outliers varied across 50 realizations. Specifically, the fraction of contaminated readouts was randomly chosen from a range of 1\% to 20\%, and the variance of the noise added to these readouts was randomly selected from a range between $\sigma^2$ and $10\sigma^2$. Also, to assess the methods in the absence of outliers, we performed an additional 5 realizations without outliers, each with a different noise variance ranging from $\sigma^2$ to $4\sigma^2$.} The four methods were compared in terms of normalized mean square error (NMSE) and structural similarity index (SSIM).


\subsection{Study II--Dynamic phantom}
\begin{figure*}[ht!]
    \begin{center}
    \includegraphics[width = 0.95\textwidth]{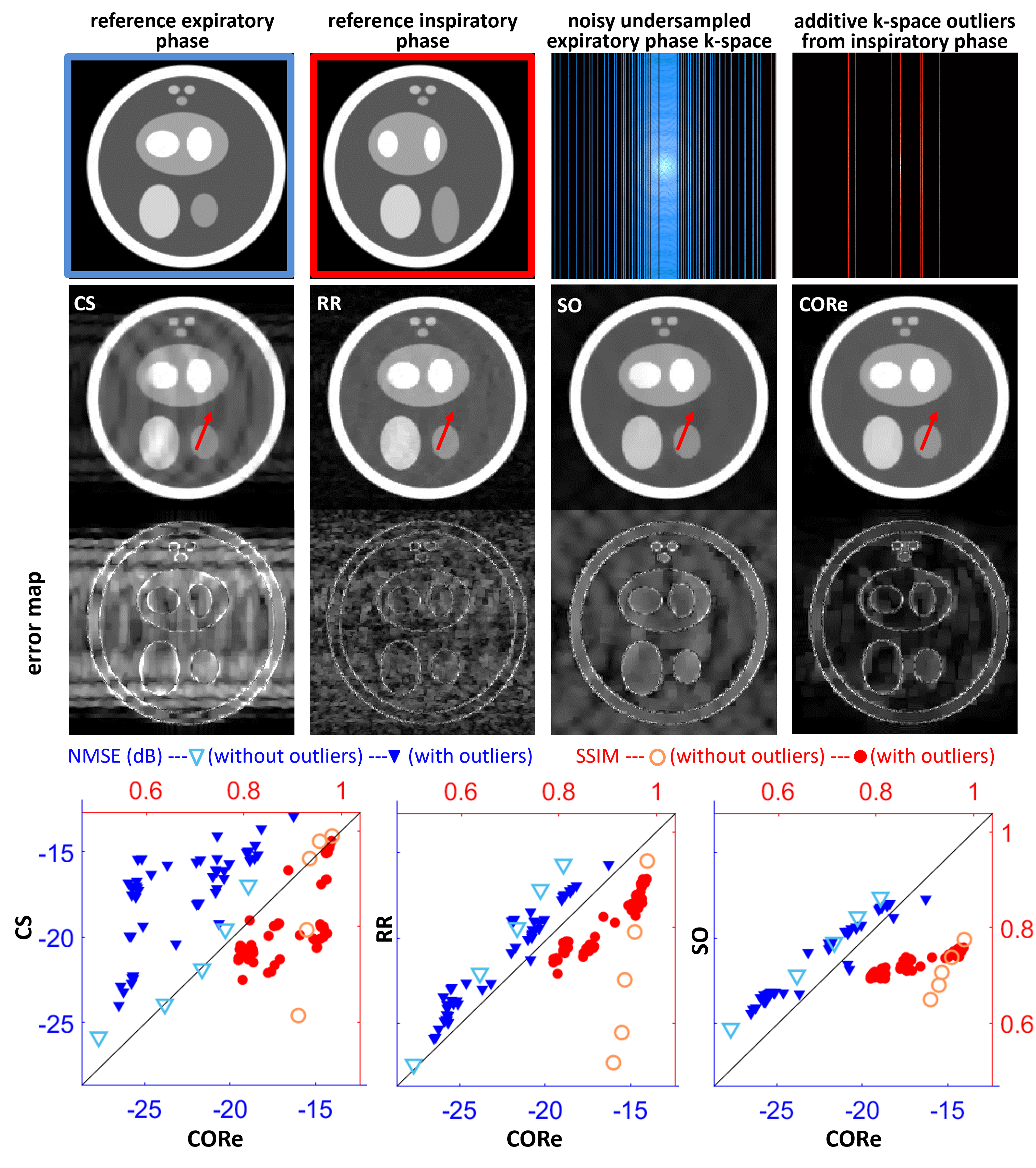}
    \end{center}
    \caption{\textr{Summary of the results from Study II. The top row (from left to right) displays the reference images from the expiratory and inspiratory phases, noisy undersampled k-space from the expiratory phase, and additive outliers from the inspiratory phase. The second row shows reconstructed images using CS, RR, SO, and \core, with artifacts highlighted by red arrows. The third row shows $3\times$ amplified error maps of the corresponding reconstructed images. The bottom row displays scatter plots for 55 realizations, comparing \core with CS, RR, and SO in terms of NMSE and SSIM, both with and without outliers.}}
    \label{fig:Figure 2}
\end{figure*}

While Study I was more generic and applicable to outliers that originated from excessive additive noise, Study II was more focused on motion artifacts. In this study, we simulated a $256\times 256$ dynamic phantom that cycles through ``inspiratory'' and ``expiratory'' states. \textr{For simplicity, cardiac motion was not included and only two respiratory motion states and single-coil data were simulated, as shown in \figref{Figure 2}.} \textr{Over a period of five respiratory cycles, a total of 260 readouts were simulated using a GRO sampling pattern similar to that employed in Study I. This resulted in a total of 130 readouts from each of the two motion states. To simulate outliers originating from imperfect data binning, we created a contaminated undersampled k-space where 90\% of the sampled readouts were derived from the expiratory state and 10\% originated from the inspiratory state.} The resulting motion-contaminated undersampled k-space, with added circularly symmetric white Gaussian noise of fixed variance $\sigma^2$, was then used to compare CS, RR, SO, and \core reconstructions using NMSE and SSIM. The simulation was repeated 50 times, each with a random realization of the outlier locations.  \textr{Similar to Study I, we considered an additional 5 realizations for a range of noise variance, without motion contamination.}

\subsection{Study III--High resolution 3D cine}

In Study III, we compared CS and \core using seven high spatial resolution 3D cine datasets collected using a self-gated, free-running sequence with a fixed acquisition time of 5 minutes using Cartesian sampling.\cite{joshi2022GRO} The field-of-view (FOV) was selected to visualize the aortic valve or to cover the whole heart. The data were collected on a clinical 1.5 T scanner (MAGNETOM Sola, Siemens Healthcare, Erlangen, Germany) and a clinical 3 T scanner (MAGNETOM Vida, Siemens Healthcare, Erlangen, Germany) equipped with 28-channel and 48-channel receive coils, respectively. Seven datasets were collected from consented subjects (age range, 23-75 years; 1 female; 4 patients, datasets \#1-\#4). Dataset \#7 was acquired under exercise stress using a supine cycle ergometer (MR Ergometer Pedal, Lode B.V., Groningen, The Netherlands). Additional scan parameters are summarized in \tabref{Table 1}. The range of acceleration rates is large due to significant differences in FOVs and spatial resolutions across the seven datasets.

\textr{Cardiac and respiratory gatings were performed using the self-gating readouts that were repeatedly acquired in the superior-inferior (SI) direction. Self-gating lines were reorganized to form a Casorati matrix. Applying two band-pass filters along the temporal dimension of the matrix, tailored to the frequency ranges of cardiac and respiratory motion, preceded the final step of PCA to extract the motion surrogate signals. From the cardiac signal, the k-space data were binned into 20 cardiac phases for all subjects. From the respiratory signal, the k-space data belonging to the expiratory phase was selected using soft gating with an efficiency of 50\%.} \cite{pruitt20214dflow} The data from the expiratory bin were used to reconstruct 3D cine images. In order to assess the quality of reconstructed images and evaluate the effectiveness of \core in reducing motion artifacts, we asked three expert readers to blindly score CS and \core reconstructions. To facilitate scoring, a 2D cine series was selected from each 3D reconstruction. Each image series was scored on a five-point Likert scale (5: excellent, 4: good, 3: adequate, 2: poor, 1: nondiagnostic) for both level of artifacts and image sharpness.

\begin{table*}[ht!]
 \centering
\scriptsize
\renewcommand{\arraystretch}{1.5}
\begin{tabular}{|c|C{2.5cm}|C{2.5cm}|C{2.5cm}|C{2.5cm}|}
\hline
\textbf{Parameter}   &\textbf{2D-PC}&\textbf{3D cine}&\textbf{Rest 4D flow}&\textbf{Stress 4D flow}\\ \hline\hline
 No. of datasets& 12 (2 patient)& 7 (4 patients)&12 (2 patients)&8 (2 patients)\\ \hline
  Field strength&$11$ at 3 T,
$1$ at 1.5 T& $5$ at 1.5 T, 
$2$ at 3 T&$11$ at 3 T,
$1$ at 1.5 T &3 T\\\hline
TE/TR (ms)      &$2.1/3.5$&$1.2-1.5/3.1-3.3$&  $2.7-2.8/4.9-5.1$ &$2.3-2.8/4.5-5.1$\\ \hline
FOV (mm)      &$347\times 260-420\times 315$&$240\times 240\times 52-440\times 440\times 144$  &   $220\times 220\times 151-300\times 300\times 151$ &$220\times 220\times 151-300\times 300\times 151$\\ \hline
Spatial Res. (mm)      &$2.0\times 2.7\times 6.0 - 2.4\times 3.3\times 6.0$&$1.3\times 1.3\times 1.0 - 2.1\times 2.3  \times 2.0$&  $2.3\times 2.3\times 2.7 - 3.1\times 3.1\times 2.7$ &$2.3\times 2.3\times 2.7 - 3.1\times 3.1\times 2.7$\\ \hline
Temporal Res. (ms)      &$35-42$&$32-50$   &  $33-54$ &$25-40$\\ \hline
Sequence     &GRE&$3\times$bSSFP, $4\times$GRE& GRE 4-point encoding  &GRE 4-point encoding  \\ \hline
Acc. rate $(R)$     &16&$7-34$ & $23-27$  &$23-25$\\ \hline
Flip angle (degrees) &$10$&$33-60$ bSSFP, $12-14$ GRE& $7-20$ &$7$\\ \hline
Acquisition time (s)&$3$ or $6$&$240-300$&  $300$ &$300$ \\\hline 
 Respiratory efficiency (\%)& real-time& 50& 50&50\\\hline 
 Self-gating frequency& $--$&  $10\times TR$ or $20\times TR$& $8\times TR$&$8\times TR$\\\hline 
VENC (cm/s)  &$--$&$--$  &  $150-200$&$200-500$\\ \hline
\end{tabular}
\caption{Imaging parameters for 2D-PC, 3D cine, 4D flow at rest, and 4D flow during exercise studies.\\\textit{Note}: Ranges indicate minimum and maximum values.}
\label{tab:Table 1}
\end{table*}

\subsection{Study IV--Rest 4D flow}
In Study IV, we compared CS and \core for free-running 4D flow imaging performed at rest. Twelve datasets (age range, 22-68 years; 4 females; 2 patients, dataset \#11 and dataset \#12) were collected using a self-gated, free-running research sequence with a fixed scan time of 5 minutes using a Cartesian sampling proposed by Pruitt et al.\cite{pruitt20214dflow} The first eleven datasets in Table 4 were acquired on a clinical 3 T scanner (MAGNETOM Vida, Siemens Healthcare, Erlangen, Germany) equipped with 48-channel receive coils. The last dataset (\#12) in \tabref{Table 4} was collected on a clinical 1.5 T scanner (MAGNETOM Sola, Siemens Healthcare, Erlangen, Germany) equipped with 28-channel receive coils. \textr{To compare net flow (ml/beat) and peak velocity (cm/s) at ascending aorta (Aao) using \core and CS, real-time 2D phase-contrast MRI (2D-PC) was used as a reference.} The imaging volume was selected to cover the whole heart and aortic arch. \tabref{Table 1} provides a summary of the additional acquisition parameters. 
The cardiac binning and soft respiratory gating were performed as described in Study III. Using \core and CS, magnitude and three velocity components were reconstructed: $\mathrm{v_{x}}$ in the superior-inferior direction, $\mathrm{v_{y}}$ in the right-left direction, and $\mathrm{v_{z}}$ in the anterior-posterior direction. After reconstruction, 4D flow images underwent background phase correction.\cite{pruitt2019bpc} Then, the images were converted to DICOM format and analyzed in CAAS (Pie Medical Imaging B.V., Maastricht, The Netherlands) for flow assessment.

\subsection{Study V--Stress 4D flow}

In Study V, we performed 4D flow imaging to evaluate the efficacy of \core in comparison to CS for mitigating motion artifacts in the presence of exercise-induced motion. Eight datasets (age range, 26-68 years; 2 females; 2 patients, datasets \#7 and \#8) were acquired during in-magnet exercise using the same research sequence as described in Study IV. All datasets were collected on a clinical 3 T scanner (MAGNETOM Vida, Siemens Healthcare, Erlangen, Germany) equipped with a supine cycle ergometer (MR Ergometer Pedal, Lode B.V., Groningen, The Netherlands). \tabref{Table 1} provides a summary of the additional acquisition parameters. The DICOM images were reconstructed using CS and \core, as described previously in Study IV. 
\textr{In the absence of a reference 2D-PC under exercise stress, we compared CS and \core by evaluating the mean and standard deviation (SD) of net flow across five transecting planes perpendicular to the flow direction at both the ascending and descending aorta, as depicted in Figure 5A. These parallel planes were positioned closely, yet away from any vessel bifurcation, maintaining an approximate distance of 6-8 mm between adjacent planes. Therefore, in the absence of motion artifacts, one would expect the net flow through the five slices to be consistent.}

\subsection{Image reconstruction}
All reconstruction methods were implemented using the ADMM algorithm.\cite{admm_1975,admm_2014} The derivation of the ADMM algorithm to solve \eqref{core} is given in Section (B) in Appendix. In each outer iteration of ADMM, a sequence of simpler subproblems is solved either in closed form or using a small number of gradient descent iterations. 
In Study I, we utilized 500 outer iterations, and the image reconstruction time with CPU processing was approximately 9 seconds for each method. For Study II, we used 250 outer iterations, and the image reconstruction time with CPU processing was approximately 30 seconds for each method. For all in vivo studies, coil compression was performed to generate 12 virtual coils for faster processing,\cite{buehrer2007array} and the coil sensitivity maps were estimated using the method by Walsh et al. \cite{walsh2000adaptive} Image reconstruction was performed using 50 outer iterations on an NVIDIA RTX 3090 GPU. The reconstruction times for both CS and \core in Studies III, IV and V were $25$, $11$ and $11$ minutes each, respectively. All images were reconstructed offline using MATLAB (Mathworks, Natick, Massachusetts).

\textr{In Study I and Study II, the hyperparameters, i.e., $\lambda_1$ for CS and RR and $\lambda_1$ and $\lambda_2$ for SO and \core, were optimized using two additional realizations, one with and one without the outliers. The optimization aimed to minimize the overall NSME (in dB) defined as $20\log_{10}\left(\frac{\| \hvec{x} - \vec{x} \|_2}{\| \vec{x} \|_2}\right)$. For the 3D cine study, the hyperparameters for CS and \core were selected based on visual inspection of a separate dataset, which was not included in the current study. For 4D flow analysis, the hyperparameters were chosen to minimize the bias from 2D-PC net flow value using another dataset not included in this study.}

\section{Results}\label{sec:res}

\subsection{Study I--Static phantom}
\figref{Figure 1} summarizes the results of Study I. The top row shows the noiseless reference image, noisy undersampled k-space, and additive outliers for one representative realization. The second and third rows show reconstructions from CS, RR, SO, and \core and their respective error maps after three-fold amplification. The bottom row shows scatter plots for 55 realizations, comparing \core with CS, RR, and SO in terms of NMSE and SSIM. The averaged NMSE and SSIM values are reported in \tabref{Table 2}, with bold values representing the best results. \textr{Statistical analysis using paired sample t-test with $\alpha=0.01$ and Bonferroni correction reveals that \core exhibits significant improvement over CS, RR, and SO in terms of both NMSE (dB) and SSIM. Figure S2 visually compares outliers rejected by SO and \core for a representative case in this study. Although this simulation study is not directly tied to motion-related outliers, it demonstrates the merit of \core for a broader application, where some of the readouts are corrupted by higher variance noise.}

\begin{table}
    \centering
    \begin{tabular}{|C{1.5cm}|C{2.2cm}|C{2.2cm}|} \hline 
         \textbf{Study I}&\textbf{NMSE (dB)}&\textbf{SSIM}\\ \hline 
         \textbf{CS}&  -20.2& 0.894\\ \hline 
         \textbf{RR}&  -24.5& 0.950\\ \hline 
         \textbf{SO}&  -23.9& 0.966\\ \hline 
         \textbf{\core}&  \textbf{-26.2}& \textbf{0.970}\\ \hline \hline 
         \textbf{Study II}& \textbf{NMSE (dB)}&\textbf{SSIM}\\ \hline
         \textbf{CS}&  -17.9& 0.809\\ \hline 
         \textbf{RR}&  -21.2& 0.796\\ \hline 
         \textbf{SO}&  -21.3& 0.726\\ \hline 
         \textbf{\core}&  \textbf{-22.7}& \textbf{0.896}\\ \hline
    \end{tabular}
    \caption{\textr{Average NMSE (dB) and SSIM results derived from 55 random draws in Study I and II. Bold values indicate the best results.}}
    \label{tab:Table 2}
\end{table}

\subsection{Study II--Dynamic phantom}
 \figref{Figure 2} summarizes the results of Study II. The top row shows the reference inspiratory and expiratory motion states of the bimodal phantom, noisy undersampled k-space from the expiratory phase, and additive motion outliers from the inspiratory phase for one representative realization. The second row shows the reconstructed images from CS, RR, SO, and \core with arrows highlighting the motion artifacts. The third row shows the respective error maps after three-fold amplification. The bottom row shows scatter plots for 55 realizations, comparing \core with CS, RR, and SO in terms of NMSE and SSIM. The averaged NMSE and SSIM values are reported in \tabref{Table 2}, with bold values representing the best results. \textr{Statistical analysis using paired sample t-test with $\alpha=0.01$ and Bonferroni correction underscores that \core exhibits significant improvement over CS, RR, and SO in terms of both NMSE (dB) and SSIM.}

\subsection{Study III--High resolution 3D cine}
\begin{figure*}[ht!]
    \begin{center}
    \includegraphics[width = 0.9\textwidth]{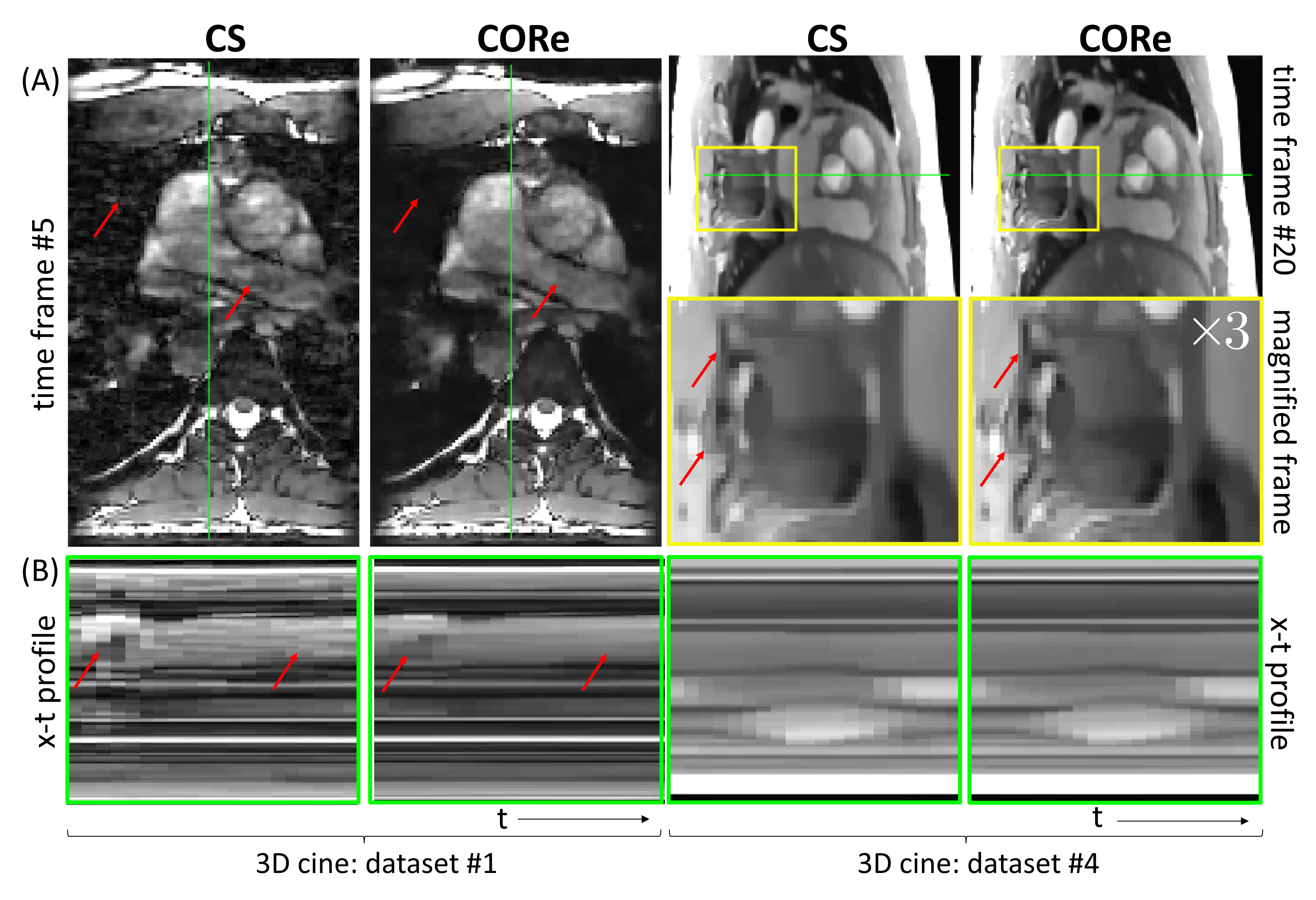}
    \end{center}
    \caption{The top row (A) shows a visual comparison of a single representative frame from two different datasets (\#1 and \#4) reconstructed using CS and \core in Study III. The example on the left highlights the reduction of motion and flow artifacts in the axial view, as indicated by the red arrows. In the example on the right, the \core image appears to preserve more detail, as emphasized by the red arrows in $3\times$ magnified frame. The x-t profiles shown in (B) consist of 20 temporal frames, plotted for the location highlighted by the green line in (A). The red arrows highlight the comparison of artifacts in CS and \core reconstructions. }
    \label{fig:Figure 3}
\end{figure*}

\textr{\tabref{Table 3} presents the averaged scores obtained from the blinded reader evaluation of the reconstructed image series, with the best scores in bold font. Our analysis compared \core and CS using a mixed-effects model with Bonferroni correction. The results indicate that \core outperforms CS in terms of artifact reduction and image sharpness, with these differences being statistically significant for $\alpha = 0.01$. \figref{Figure 3}A provides a visual comparison of the 3D cine reconstructions from CS and \core, showing a representative frame from two different datasets (\#1 and \#4). The arrows in the figure highlight visual differences between CS and \core in terms of artifacts or image sharpness. \figref{Figure 3}B compares x-t profiles for locations highlighted with green lines in \figref{Figure 3}A. Ancillary Video S1 shows CS and \core reconstructed cine image series for dataset \#1.}

\begin{table}[ht!]
 \centering
\scriptsize
\renewcommand{\arraystretch}{1.5}
\begin{tabular}{|C{0.8cm}|l|l|l|l|}
\hline
\multirow{2}{0.8cm}{{\textbf{Data-set}}} &\multicolumn{2}{|c|}{\textbf{Artifacts}} &\multicolumn{2}{|c|}{\textbf{Sharpness}} \\\cline{2-5}
& \multicolumn{1}{|c|}{{\textbf{CS}}} & \multicolumn{1}{|c|}{{\textbf{\core}}} & \multicolumn{1}{|c|}{{\textbf{CS}}} & \multicolumn{1}{|c|}{{\textbf{\core}}} \\  \hline\hline
 \#1&   $2.7 \pm 0.6$&  $\textbf{4.0} \pm 0.7$& $3.3 \pm 1.2$& $\textbf{3.7} \pm 1.5$\\
\#2&  $4.3 \pm 0.6$&  $\textbf{4.7} \pm 0.6$& $3.7 \pm 0.6$&$3.7 \pm 0.6$\\
\#3&  $3.3 \pm 1.2$& $\textbf{4.0} \pm 1.0$& $2.7 \pm 1.2$&$2.7 \pm 1.2$\\
\#4&   $4.3 \pm 1.2$& $\textbf{4.7} \pm 0.6$& $3.3 \pm 1.2$& $3.3 \pm 1.2$\\
\#5&   $3.3 \pm 1.2$&$\textbf{3.7} \pm 1.5$& $1.7 \pm 1.2$& $\textbf{2.0} \pm 1.0$\\
\#6& $2.7 \pm 1.5$& $\textbf{4.0} \pm 1.0$& $2.0 \pm 1.0$&$\textbf{3.0} \pm 1.0$\\
\#7& $2.3 \pm 1.2$& $\textbf{3.7} \pm 1.5$& $2.0 \pm 1.0$&$\textbf{2.7} \pm 1.5$\\\hline
{\textbf{Avg.}} &   $3.3$& $\textbf{4.1}$&$2.7$& $\textbf{3.0}$\\

\hline
\end{tabular}
\caption{Results of a blinded reader study conducted on seven 3D cine datasets. Each score represents an average$\pm$SD from three CMR expert readers. Bold values indicate the better scores for each criterion.}
\label{tab:Table 3}
\end{table}

\subsection{Study IV--Rest 4D flow}
\begin{figure*}[!ht]
    \begin{center}
    \includegraphics[width = 0.9\textwidth]{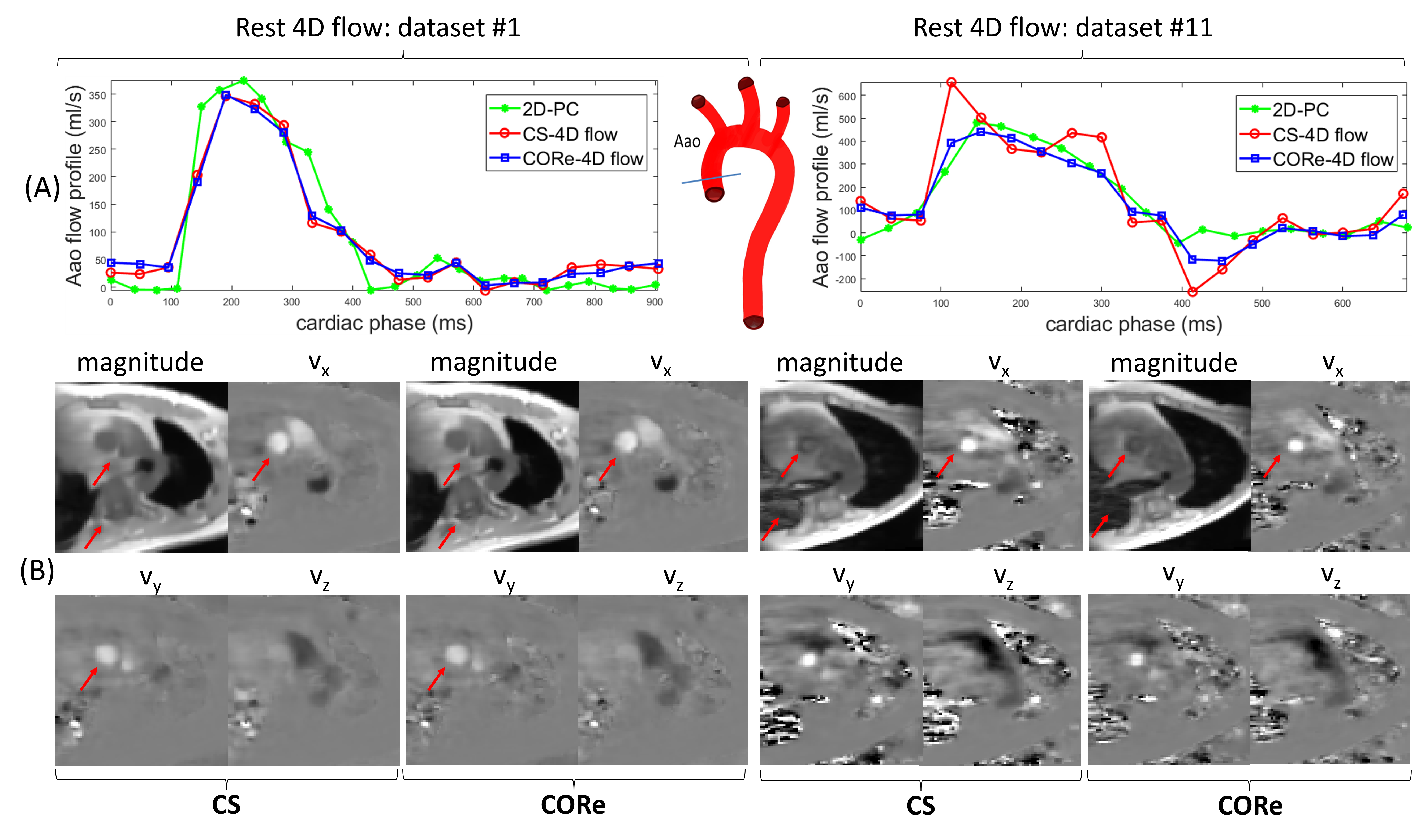}
    \end{center}
    \caption{Qualitative comparison of CS and \core reconstructions using representative aortic flow profiles, magnitude, and velocity components of rest 4D flow images in Study IV. The center of the top row (A) shows a 2D plane transecting ascending aorta (Aao) used for comparing Aao flow profiles measured from CS and \core with 2D-PC reference. (A) shows representative volumetric flow rate profiles measured using 4D flow reconstructions and 2D-PC reference. (B) presents representative magnitude image, velocity component in the superior-inferior direction ($\mathrm{v_{x}}$), velocity component in the right-left direction ($\mathrm{v_{y}}$), and velocity component in the anterior-posterior direction ($\mathrm{v_{z}}$) from CS and \core. A single frame at systole (peak flow) is shown for the 2D plane defined in (A). The VENC values for dataset \#1 and dataset \#11 were $150$ cm/s and $200$ cm/s, respectively. For better visual comparison, the velocity components are amplified by a factor of 2. The red arrows highlight areas where CS exhibits artifacts or blurring.}
    \label{fig:Figure 4}
\end{figure*}

\tabref{Table 4} compares net flow quantification (ml/beat) and peak velocities (cm/s) from CS and \core with 2D-PC serving as reference. The flow measurements were performed at an Aao plane depicted in \figref{Figure 4}A. The bottom row in the table shows the mean absolute error (MAE) of CS and \core values from the 2D-PC reference. A paired sample t-test, with Bonferroni correction, indicates that the differences of both CS and \core from 2D-PC are statistically insignificant at a significance level of $\alpha=0.01$. \figref{Figure 4}A provides a visual comparison of two representative flow rate profiles from CS and \core. \figref{Figure 4}B shows the magnitude and three velocity components corresponding to datasets \#1 (left) and \#11 (right). A single frame at systole (peak flow) is shown from a 2D slice transecting Aao as depicted in \figref{Figure 4}A. The arrows highlight areas where CS exhibits motion artifacts or blurring. Ancillary Video S2 displays CS and \core reconstructed cine image series for magnitude, $\mathrm{v_{x}}$, $\mathrm{v_{y}}$, and $\mathrm{v_{z}}$ components of dataset \#11. Peak velocity assessment from dataset \#12 was not feasible due to localized signal loss associated with a prosthetic aortic valve.

\subsection{Study V--Stress 4D flow}
\begin{figure*}[!ht]
    \begin{center}
    \includegraphics[width = 0.9\textwidth]{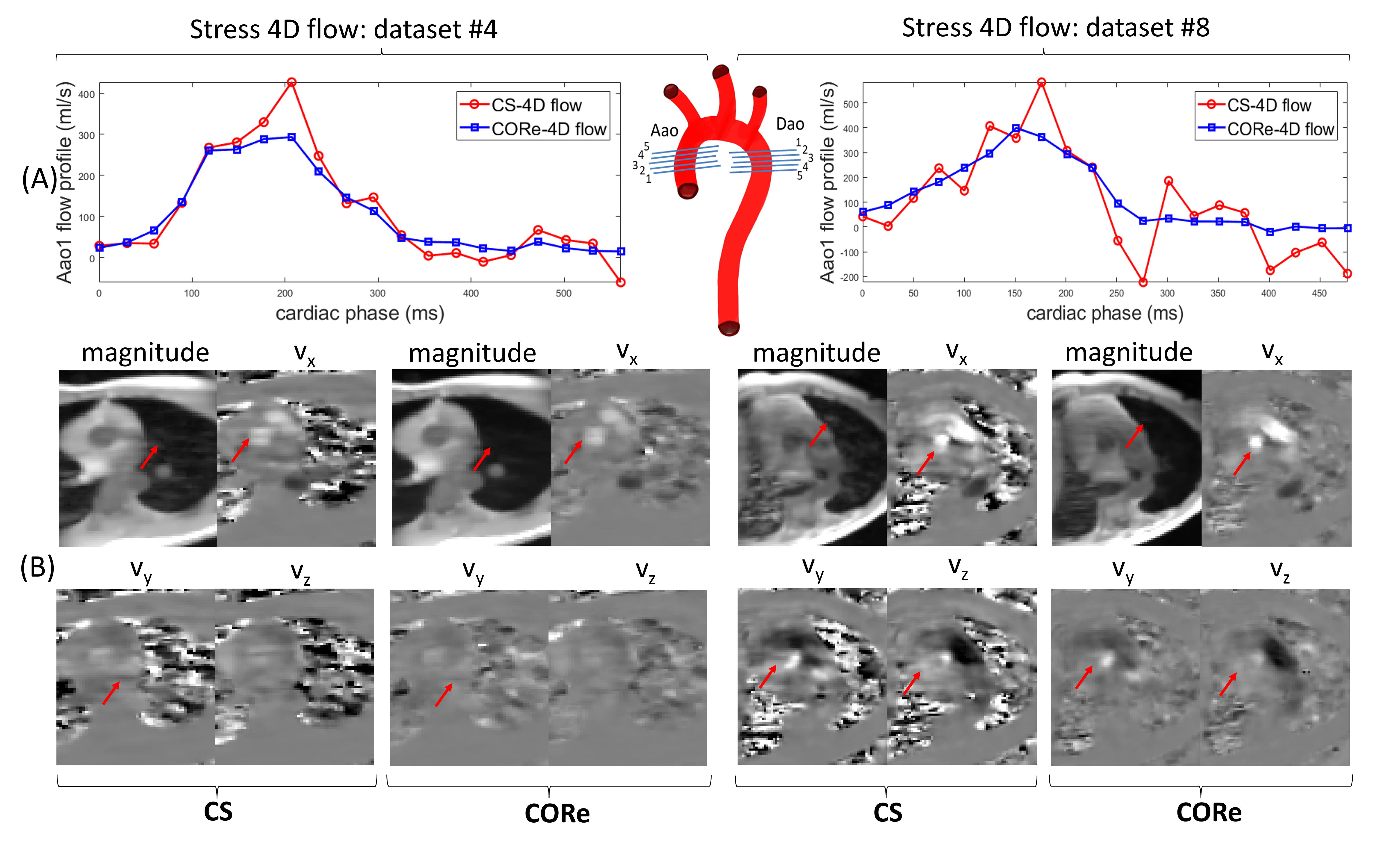}
    \end{center}
    \caption{Qualitative comparison of CS and \core reconstructions using representative aortic flow profiles, magnitude and velocity components of stress 4D flow images in Study V. The center of top row (A) shows five transecting planes defined at ascending aorta (Aao) and descending aorta (Dao) for stress 4D flow analysis in Study V. (A) shows representative volumetric flow rate profiles at Aao plane 1 measured using 4D flow reconstructions. (B) presents representative magnitude image, velocity component in the superior-inferior direction ($\mathrm{v_{x}}$), velocity component in the right-left direction ($\mathrm{v_{y}}$), and velocity component in the anterior-posterior direction ($\mathrm{v_{z}}$) from CS and \core are shown at systole (peak flow) for the 2D plane 1. The VENC values for dataset \#4 and dataset \#8 were $250$ cm/s and $200$ cm/s, respectively. For better visual comparison, the phase contrast images are amplified by a factor of 2. The red arrows highlight areas where CS exhibits artifacts or blurring.}
    \label{fig:Figure 5}
\end{figure*}

Table 5 compares CS and \core for mean$\pm$SD values of net flow quantification (ml/beat) across the Aao and Dao planes depicted in \figref{Figure 5}A. The lower SD values are highlighted with bold font. For $\alpha = 0.01$, a paired sample t-test indicates a statistically significant difference in standard deviations between \core and CS. \figref{Figure 5}A provides a visual comparison of two representative flow rate profiles from CS and \core at Aao plane 1. \figref{Figure 5}B shows the magnitude and three velocity components corresponding to datasets \#4 (left) and \#8 (right). A single frame at systole (peak flow) is shown from Aao plane 1. The arrows highlight areas where CS exhibits motion artifacts or blurring. Ancillary Video S3 displays CS and \core reconstructed cine image series for magnitude, $\mathrm{v_{x}}$, $\mathrm{v_{y}}$, and $\mathrm{v_{z}}$ components of dataset \#8. The Video consists of 20 frames from one cardiac cycle, played four times at a frame rate of 10 frames per second.

\section{Discussion}\label{sec:dis}
Free-running self-gated volumetric CMR is prone to artifacts due to the presence of outliers in binned k-space data. To address this issue, we propose a novel reconstruction method, \core (\eqref{core}), which effectively integrates outlier rejection into compressive reconstruction. \core leverages the group sparse behavior of outliers in k-space to separate them from properly binned measurements, resulting in reduced artifacts. \\
We first conduct two simulation studies to quantitatively compare \core with CS (\eqref{map1}), RR (\eqref{map2}), and SO (\eqref{l2l1v}). In Study I, we simulate noisy undersampled k-space from a static phantom and contaminate a fraction of the readouts with stronger additive complex noise to mimic outliers. \textr{The reconstruction methods were assessed with and without the presence of outliers in the data using NMSE and SSIM metrics. The results from 55 realizations demonstrate that CS reconstructions suffer significant quality degradation in the presence of outliers. While RR and SO methods exhibit noticeable improvement over CS, \core outperforms them, achieving the best average NMSE and SSIM values, as presented in \tabref{Table 2}.} In Study II, undersampled k-space readouts are combined from two different motion states of a dynamic phantom to mimic free-running CMR acquisition with imperfect binning. Again, in terms of NMSE and SSIM averaged over 55 realizations, \core outperforms CS, RR, and SO, as reported in \tabref{Table 2}. The error maps also highlight residual motion artifacts in CS, RR and SO that are not visible in \core. Figure S2 highlights the benefit of group sparsity used in \core over unstructured sparsity used in SO. This figure shows the error maps, \hvec{v}, represented in k-space for one of the examples in Study I. Compared to \core, SO is unable to reject the entire readouts that are outliers.\\
In Study III, we evaluate \core for reconstructing high-resolution 3D cine images. The comparison is made with CS, which is a common choice for FRV reconstruction. We did not include RR and SO as these methods have not been previously proposed or validated for MRI reconstruction. The reconstructed images from \core and CS were blindly scored by three expert readers on the criteria of artifacts and sharpness. The results from the reader study (presented in \tabref{Table 3}) show that \core not only outperforms CS in terms of artifacts but also provides marginal but consistent improvement in image sharpness. The suppression of artifact by \core is evident in \figref{Figure 3} (left) and the corresponding Ancillary Video S1. The improvement in sharpness by \core is evident in \figref{Figure 3} (right). The simultaneous reduction in artifacts and improvement in sharpness by \core indicate that the relatively inferior performance of CS cannot be attributed to suboptimal selection of the regularization strength, $\lambda_1$. It is well known that the regularization strength provides a trade-off between artifact (or noise) suppression and image sharpness.\cite{khare2012lambda} For CS, increasing $\lambda_1$ to suppress artifacts would lead to further image blurring, and decreasing $\lambda_1$ to improve image sharpness would further amplify image artifacts.\\  
In Study IV, we evaluate CS and \core for the reconstruction of 4D flow data acquired at rest, using 2D-PC as a reference. In particular, we quantify the net flow (ml/beat) and peak velocity (cm/s) across a plane transecting the ascending aorta (Aao) as demonstrated in \figref{Figure 4}A. The flow quantification results presented in \tabref{Table 4} demonstrate that both CS and \core reconstructions exhibit agreement with 2D-PC measurements. The MAE values indicate that CS and \core produce comparable results, with \core showing a marginal reduction in bias. However, in a few datasets, \core underestimates the net flow and CS tends to overestimate in comparison to the reference. For (patient) dataset \#11, CS demonstrates a significant overestimation of net flow when compared to the reference, evident from the oscillations observed in the CS reconstructed flow rate profile (ml/s) depicted in \figref{Figure 4}A. This jagged flow behavior can be attributed to the potential incorrect binning of k-space data, likely caused by bulk motion by the subject during the scan. This interpretation is supported by the cardiac and respiratory signals displayed in Figure S3A. On the contrary, the flow profile from \core reconstruction for the same dataset is consistent with the 2D-PC flow profile, demonstrating \core's ability to mitigate motion-induced outliers in k-space data. Additionally, the visual comparison of magnitude and velocity components in \figref{Figure 4}B and Ancillary Video S2 shows that \core reconstruction effectively reduces motion artifacts while preserving finer details compared to CS. \\
In Study V, we assess CS and \core for the reconstruction of 4D flow data acquired during exercise stress. Since the self-gating signal is less reliable under exercise stress, the k-space data collected under exercise stress are more susceptible to motion outliers. We compare the consistency of flow across multiple planes transecting Aao and Dao regions as depicted in \figref{Figure 5}A. The flow quantification results presented in \tabref{Table 5} show that \core demonstrates considerably lower variation in net flow values across the planes compared to CS. The representative flow profiles shown in \figref{Figure 5}A illustrate that flow profiles obtained from \core images appear smoother and more physiologically realistic. In contrast, CS flow profiles exhibit more jagged behavior, potentially due to uncompensated motion outliers. This interpretation is supported by the cardiac and respiratory signals shown for dataset \#8 in Figure S3B. The improvement offered by \core in mitigating motion artifacts can be clearly observed in the representative 4D flow images presented in \figref{Figure 5}B and Ancillary Video S3. This demonstrates that the impact of \core becomes even more pronounced in stress 4D flow imaging.\\
Although in this work \core is only applied to Cartesian sampling, it is equally applicable to non-Cartesian sampling and applications other than cine and flow. Also, \core makes no assumption about the genesis of outlier readouts and can potentially suppress outliers originating from sources other than motion, such as RF interference and flow artifacts. Finally, the central concept of \core can be integrated with deep-learning-based reconstruction methods\cite{aggarwal2018modl} and motion correction-based reconstruction methods\cite{dwight2014motioncorr} by modifying the data-consistency term. \\
\core implementation using the ADMM algorithm does not significantly increase computational cost over CS. The reconstruction times for \core and CS are comparable. Still, this work has several limitations. First, compared to CS, \core introduces another tuning parameter, $\lambda_2$. In all our studies, this parameter was optimized using an additional dataset that was not included in the comparison. For applications, where imaging parameters vary widely across subjects, using one value of $\lambda_2$ may not be optimal. Employing a smaller value for $\lambda_2$ than the optimal choice may risk discarding valid k-space measurements. Conversely, selecting a larger value for $\lambda_2$ may diminish the advantage of \core over CS in terms of outlier suppression. Second, \core relies on the assumption that coil sensitivity maps, which are estimated from the measured k-space that includes outliers, are of high quality. In cases where this assumption breaks down, the final quality of \core reconstruction may not be adequate. A possible solution to this problem is to estimate the coil sensitivity maps again after a preliminary \core reconstruction and then perform a final \core reconstruction using the updated maps. This remedy, however, would come at the cost of increased reconstruction time. \textr{Third, the in vivo studies include a limited number of subjects, with evaluation relying on subjective assessment of image quality in Study III, quantification of only two hemodynamic parameters in Study IV, and internal consistency in Study V. Since artifacts can significantly impact quantification of advanced hemodynamic parameters, including wall shear stress and pressure gradients, \core is expected to improve the reliability of such parameters. Future studies will include larger in vivo studies, quantitative comparison of cardiac function to 2D imaging, and assessment of advanced hemodynamic parameters.} 

\begin{table*}[!ht]
 \centering
\scriptsize
\renewcommand{\arraystretch}{1.5}
    \begin{tabular}{|C{0.8cm}|C{1.2cm}|C{1.2cm}|C{1.2cm}|C{1.2cm}|C{1.2cm}|C{1.2cm}|} 
    
    \hline 
        \multirow{2}{1cm}{{\textbf{Data-set}}}&  \multicolumn{3}{c}{\textbf{Net flow  (ml/beat)}} &  \multicolumn{3}{|c|}{\textbf{Peak velocity  (cm/s)}} \\ \cline{2-7}
         &  \textbf{2D-PC}&  \textbf{CS}& \textbf{\core} &  \textbf{2D-PC} &  \textbf{CS}& \textbf{\core} \\ \hline\hline 
        \#1 & 86.9 & 84.4 & 85.0 & 75 & 84 & 83 \\\hline
        \#2 & 74.9 & 72.5 & 71.4 & 77 & 89 & 89 \\\hline
        \#3 & 79.2 & 76.7 & 75.5 & 84 & 106 & 98 \\\hline
        \#4 & 48.7 & 47.2 & 47.8 & 140 & 100 & 109 \\\hline
        \#5 & 74.6 & 71.4 & 67.4 & 114 & 119 & 116 \\\hline
        \#6 & 83.7 & 72.1 & 71.7 & 108 & 106 & 107 \\\hline
        \#7 & 57.8 & 59.8 & 57.6 & 93 & 103 & 109 \\\hline
        \#8 & 64.8 & 70.6 & 53.4 & 74 & 125 & 103 \\\hline
        \#9 & 72.4 & 72.4 & 66.9 & 94 & 104 & 97 \\\hline
        \#10 & 76.0 & 72.6 & 77.2 & 75 & 88 & 85 \\\hline
        \#11 & 94.7 & 111.9 & 93.6 & 180 & 188 & 173 \\\hline
        \#12 & 35.9 & 37.9 & 37.3 & $--$ & $--$ & $--$ \\\hline
    $--$& \textbf{MAE}& 4.5&\textbf{4.2} & \textbf{MAE}& 16&\textbf{12}\\ \hline
    \end{tabular}
    \caption{\textr{Comparison of net flow (ml/beat) and peak velocity (cm/s) in Study IV at Aao plane shown in \figref{Figure 4}A, from CS and \core reconstructions of rest 4D flow, and 2D-PC reference. The last row indicates the mean absolute error (MAE) of 4D flow values with 2D-PC reference.}}
    \label{tab:Table 4}
\end{table*}

\begin{table}[!ht]
 \centering
\scriptsize
\renewcommand{\arraystretch}{1.5}
\begin{tabular}{|cc|p{2.3cm}|p{2.7cm}|}
\hline
\multicolumn{2}{|c|}{{\textbf{Dataset}}}& {\textbf{CS (ml/beat)}} & {\textbf{\core (ml/beat)}} \\  \hline\hline
\multirow{2}{*}{\#1}&Aao & $81.0\pm 2.6$& $60.3 \pm \textbf{0.4}$\\
 &Dao & $43.1 \pm 10.2$& $33.7 \pm \textbf{1.1}$\\\hline
\multirow{2}{*}{\#2}&Aao& $ 90.0\pm 4.8$&$76.7 \pm \textbf{2.1}$\\
 &Dao& $61.2 \pm 5.6$&$47.8 \pm \textbf{0.8}$\\\hline
\multirow{2}{*}{\#3}&Aao&$81.4 \pm 2.1$& $49.1 \pm \textbf{0.6}$  \\
 &Dao& $50.6 \pm 5.7$&$46.6 \pm \textbf{0.7}$\\\hline
\multirow{2}{*}{\#4}&Aao& $69.5 \pm 6.3$& $63.0 \pm \textbf{2.0}$\\
 &Dao& $60.8 \pm 3.5$& $44.6 \pm \textbf{0.8}$\\\hline
\multirow{2}{*}{\#5}&Aao& $59.5 \pm 8.4$& $71.1 \pm \textbf{1.6}$\\
 &Dao& $56.6 \pm 6.1$& $53.1 \pm \textbf{0.9}$\\\hline
 \multirow{2}{*}{\#6}& Aao& $67.4 \pm 2.3$&$59.9 \pm \textbf{1.1}$
\\
 & Dao& $58.4 \pm 0.5$&$57.4 \pm 0.5$\\\hline
  \multirow{2}{*}{\#7}& Aao& $ 245.8\pm 6.4$&$178.1 \pm \textbf{2.8}$
\\
 & Dao& $88.1 \pm 9.1$&$72.3 \pm \textbf{0.5}$\\\hline
  \multirow{2}{*}{\#8}& Aao & $73.0 \pm 14.8$&$62.2 \pm \textbf{0.8}$\\
 & Dao & $59.9 \pm \textbf{0.6}$&$55.1 \pm0.8$\\\hline
  \textbf{Avg.}& Aao & $6.0$&$\textbf{1.4}$\\
 \textbf{SD}& Dao & $5.1$&$\textbf{0.8}$\\\hline
\end{tabular}
\caption{Comparison of flow quantification from CS and \core reconstructions of stress 4D flow in Study V. The values represent the mean$\pm$SD net flow across Aao and Dao planes, as defined in \figref{Figure 5}A. Bold values indicate the lower standard deviation. The last row indicates the average standard deviation across Aao and Dao in CS and \core reconstructions in Study V.}
\label{tab:Table 5}
\end{table}

\section{Conclusion}\label{sec:con}
The proposed method, \core, integrates outlier rejection into the compressive reconstruction. The results from the simulation studies show that \core outperforms standard CS and other robust regression models in terms of image quality. The results from the human subject studies demonstrate that \core is more effective than CS in suppressing motion artifacts due to imperfect binning. \core's ability to suppress motion artifacts makes it an attractive candidate for free-running volumetric CMR.

 \section*{Acknowledgments}
We thank Xiaokui Mo for
her assistance with the statistical analysis.%

\section*{Data Availability Statement}
The MATLAB implementation of \core and the competing methods for all studies, along with representative self-gated k-space datasets for in vivo studies, is available on GitHub at {\href{https://github.com/OSU-MR/motion-robust-CMR}{https://github.com/OSU-MR/motion-robust-CMR}}.



\subsection*{Financial disclosure}
The authors have no relevant financial disclosures.

\subsection*{Conflict of interest}
The authors declare no conflict of interests.

\bibliography{root}%
\vfill\pagebreak

\section*{Ancillary Files}\label{sec:supp}
The following ancillary files are available as part of the online article:





\vskip\baselineskip\noindent\noindent
\textbf{Ancillary Video S1.}
Comparison of representative cine image series from dataset \#1 reconstructed using CS and \core in Study III. Images consist of 20 frames from one cardiac cycle played four times at 10 frames per second. The red arrows highlight the reduction in motion and flow artifacts using \core compared to CS. 

\noindent
\textbf{Ancillary Video S2.}
Comparison of representative magnitude and velocity components of 4D flow images of dataset \#11 acquired at rest and reconstructed using CS and \core in Study IV. Image series consists of 20 cardiac frames played four times at 10 frames per second. The red arrows highlight artifacts that are better suppressed in \core reconstruction.  

\noindent
\textbf{Ancillary Video S3.}
Comparison of representative magnitude and velocity components of 4D flow images of dataset \#8 acquired under exercise stress and reconstructed using CS and \core in Study V. Image series consists of 20 cardiac frames played four times at 10 frames per second. The red arrows highlight areas where \core is more effective in suppressing motion artifacts.



\onecolumn
\appendix
\section{-- Bayesian Perspective of \core}
\label{app:bayesian}
Consider model in Equation (4), i.e., 
\begin{equation}
\vec{y} = \vec{A}\vec{x} + \vec{v} + \vec{\omega}.
\end{equation}
Now, we assume that $\vec{x}$, $\vec{v}$, and $\vec{\omega}$ are realizations of random variables $\vec{X}$, $\vec{V}$, and $\vec{\Omega}$ defined by probability density functions $p(\vec{x}) \propto \exp(-\mathcal{R}(\vec{x}))$, $p(\vec{v}) \propto \exp(-\lambda_2\|\vec{v}\|_1)$, and $p(\vec{\omega}) \propto \exp(-\|\vec{w}\|/\sigma^2)$, respectively. With these assumptions, it is easy to see that $\vec{y}$ is a realization of a random variable $\vec{Y}$ with $p(\vec{y}) \propto \exp(-\|\vec{y}-(\vec{A}\vec{x}+\vec{v})\|/\sigma^2)$. By Bayes' rule, the posterior probability $p(\vec{x},\vec{v}\,|\,\vec{y})$ is given by
\begin{align}
    p(\vec{x},\vec{v}\,|\,\vec{y}) = \frac{p(\vec{y}\,|\,\vec{x},\vec{v})p(\vec{x},\vec{v})}{p(\vec{y})}.
    \label{eq:post}
\end{align}
By ignoring the denominator in \eqref{post} as it does not depend on $\vec{x}$ or $\vec{v}$ and assuming $\vec{X}$ and $\vec{V}$ are independent, we get
\begin{align}
    p(\vec{x},\vec{v}\,|\,\vec{y}) \propto p(\vec{y}\,|\,\vec{x},\vec{v})p(\vec{x})p(\vec{v}).
    \label{eq:post2}
\end{align}
By taking the negative $\log$ of \eqref{post2} and minimizing it with respect to $\vec{x}$ and $\vec{v}$, we get
\begin{equation}
\hvec{x}_{\text{SO}} =  \argmin_{\vec{x},\vec{v}}\frac{1}{\sigma^2}\|\vec{Ax}-\vec{y}+\vec{v}\|_2^2 + \mathcal{R}(\vec{x}) + \lambda_2\|\vec{v}\|_1,
\end{equation}
which is the same as Equation (5). It also easy to see that by replacing $p(\vec{v}) \propto \exp(-\lambda_2\|\vec{v}\|_1)$ with $p(\vec{v}) \propto \exp(-\lambda_2\|g(\vec{v})\|_1)$, we arrive at the proposed \core formulation in Equation (6). Therefore, $\hvec{x}_\text{SO}$ and $\hvec{x}_\text{CO}$ can be viewed as MAP estimates under (i) circularly symmetric white Gaussian noise, (ii) sparse or group sparse prior on $\vec{V}$, (iii) prior $\mathcal{R}(\vec{x})$ on $\vec{X}$, and (iv) statistical independence between $\vec{X}$ and $\vec{V}$.

\section{-- ADMM Algorithm for \core}
\label{app:admm}
Reconsider the optimization problem in Equation (6) with $\mathcal{R}(\vec{x})=\lambda_1\|\vec{\Psi}\vec{x}\|_1$, i.e., 
\begin{equation}
\min_{\vec{x}, \vec{v}} \frac{1}{\sigma^2}\|\vec{A}\vec{x} - \vec{y} + \vec{v} \|_2^2 + \lambda_1\|\vec{\Psi}\vec{x}\|_1 + \lambda_2\|g(\vec{v})\|_{1},
\label{eq:core-repeat}
\end{equation}
where $g(\vec{v}) \triangleq [\Tilde{v}_1, \Tilde{v}_2,\dots,\Tilde{v}_J] \tran \in \Real^{J\times 1}$ with $\Tilde{v}_j = \|\vec{v}_j\|_2 \in \Real$. Here, $\vec{v}_j\in \Complex^{K\times 1}$ represents samples in the $j^\text{th}$ readout. See Figure S1 for a graphical depiction of the notation.

\begin{figure*}[h!]
    \begin{center}
    \includegraphics[width = 0.5\textwidth]{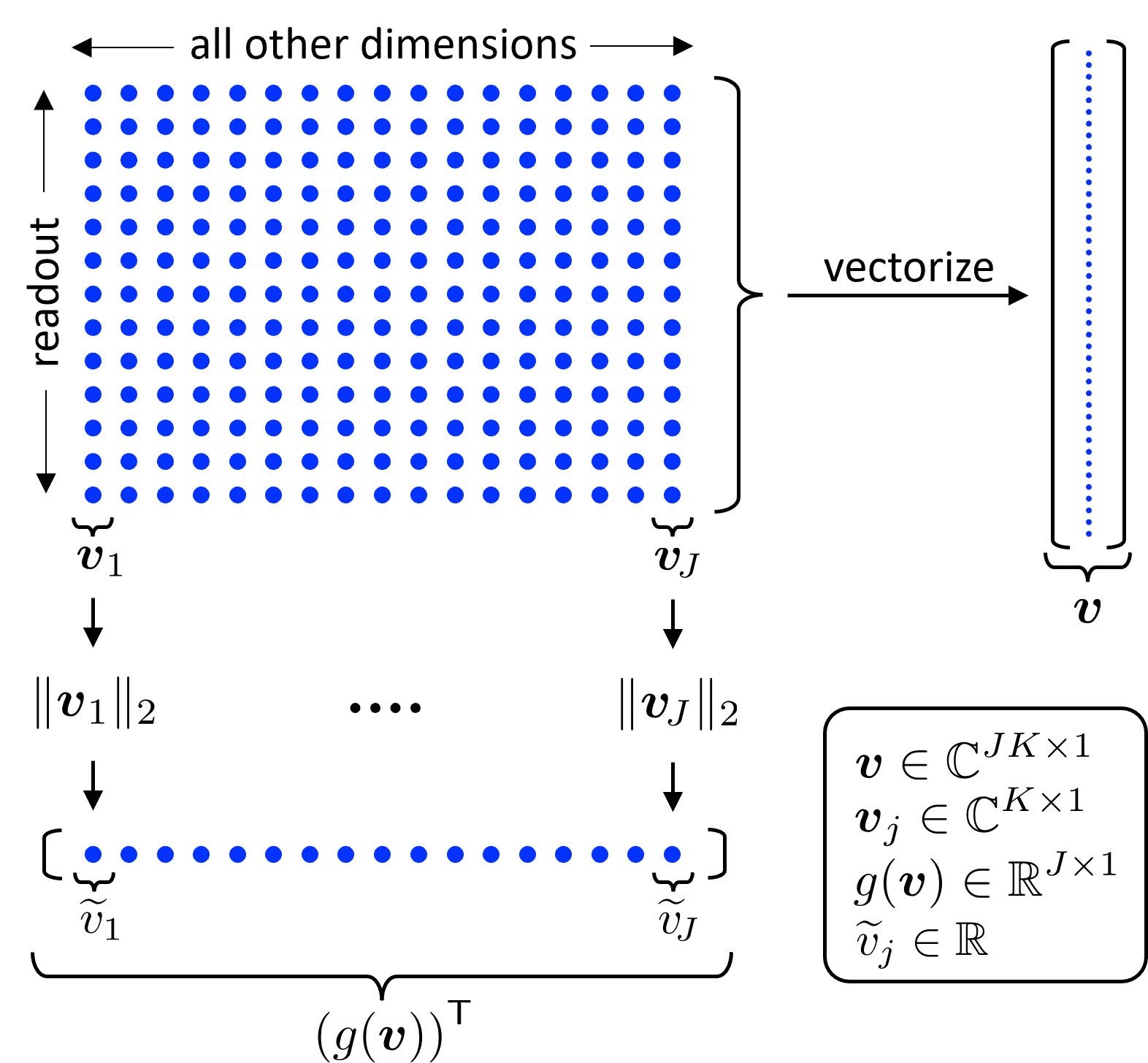}
    \end{center}
    \captionsetup{labelformat=empty}
    \caption{\textbf{FIGURE S1} Graphical depiction of the notation used to represent outliers. Here, $\vec{v}$ is the vector version of all outliers, and $g(\vec{v})$ is computed from $\vec{v}$, such that the $j^{\text{th}}$ entry in $g(\vec{v})$ represents the $\ell_2$-norm for the $j^{\text{th}}$ readout.}
    \label{fig:layout}
\end{figure*}

By $\vec{w}_1 \triangleq \vec{\Psi}\vec{x}$ and $\vec{w}_2 \triangleq g(\vec{v})$, the problem in \eqref{core-repeat} can be equivalently written as
\begin{equation}
\min_{\vec{x}, \vec{v}} \frac{1}{\sigma^2}\|\vec{A}\vec{x} - \vec{y} + \vec{v}\|_2^2 + \lambda_1\|\vec{w}_1\|_1 + \lambda_2\|\vec{w}_2\|_{1}~~~\text{subject to}~~~\vec{\Psi}\vec{x} = \vec{w}_1,~g(\vec{v}) = \vec{w}_2.
\label{eq:core-cons}
\end{equation}
The augmented Lagrangian for this problem can be expressed as
\begin{align}
    \mathcal{L}_{\mu_1, \mu_2}(\vec{x}, \vec{v}, \vec{w}_1, \vec{w}_2, \vec{\rho}_1, \vec{\rho}_2) = &~\frac{1}{\sigma^2}\|\vec{A}\vec{x} - \vec{y} + \vec{v}\|_2^2 + \lambda_1\|\vec{w}_1\|_1 + \lambda_2\|\vec{w}_2\|_{1} \nonumber \\ 
    &+ \real\big\{\vec{\rho}\herm_1(\vec{\Psi}\vec{x} - \vec{w}_1)\big\} + \frac{\mu_1}{2}\|\vec{\Psi}\vec{x} - \vec{w}_1\|_2^2 \nonumber \\ 
    &+ \real\big\{\vec{\rho}\herm_2(g(\vec{v})) - \vec{w}_2)\big\} + \frac{\mu_2}{2}\|g(\vec{v}) - \vec{w}_2\|_2^2.
    \label{eq:core-lag1}
\end{align}
By $\vec{u}_1 \triangleq \vec{\rho}_1/\mu_1$ and $\vec{u}_2 \triangleq \vec{\rho}_2/\mu_2$, we can simplify the expression in \eqref{core-lag1} to
\begin{align}
    \mathcal{L}_{\mu_1, \mu_2}(\vec{x}, \vec{v}, \vec{w}_1, \vec{w}_2, \vec{u}_1, \vec{u}_2) = &~\frac{1}{\sigma^2}\|\vec{A}\vec{x} - \vec{y} + \vec{v}\|_2^2 + \lambda_1\|\vec{w}_1\|_1 + \lambda_2\|\vec{w}_2\|_{1} \nonumber \\ 
    &+ \frac{\mu_1}{2}\|\vec{\Psi}\vec{x} - \vec{w}_1 + \vec{u}_1 \|_2^2 + \frac{\mu_1}{2}\|\vec{u}_1\|_2^2 \nonumber \\ 
    &+ \frac{\mu_2}{2}\|g(\vec{v}) - \vec{w}_2 + \vec{u}_2 \|_2^2 + \frac{\mu_2}{2}\|\vec{u}_2\|_2^2.
    \label{eq:core-lag2}
\end{align}
Now, the primal problem in \eqref{core-cons} can be solved by this minimax problem
\begin{equation}
    \min_{\vec{x},\vec{v},\vec{w}_1,\vec{w}_2} ~ \max_{\vec{u}_1, \vec{u}_2} \big\{\mathcal{L}_{\mu_1, \mu_2}(\vec{x}, \vec{v}, \vec{w}_1, \vec{w}_2, \vec{u}_1, \vec{u}_2)\big\},
    \label{eq:lag3}
\end{equation}
which leads to \algref{core} that tackles the problem presented in \eqref{lag3} by breaking it down into simpler subproblems that update one variable at a time.

\begin{algorithm}[h]
\onehalfspacing
  \caption{ADMM for CORe}
  \label{alg:core}
  \begin{algorithmic}[1]
    \REQUIRE $\mu_1>0$, $\mu_2 >0$, $\lambda_1>0$, $\lambda_2 >0$, $\sigma$, $\vec{A}$, $\vec{y}$, $\vec{x}^{(0)}=\vec{A}\herm\vec{y}$, $\vec{v}^{(0)} = \vec{0}$, $\vec{w}_1^{(0)} = \vec{0}$, $\vec{w}_2^{(0)} = \vec{0}$, $\vec{u}_1^{(0)} = \vec{0}$, $\vec{u}_2^{(0)} = \vec{0}$
	\FOR{$t=1,2,3,\dots, T$}
	\STATE{$\vec{x}^{(t)} = \argmin_{\vec{x}}\frac{1}{\sigma^2}\|\vec{A}\vec{x} - \vec{y} + \vec{v}^{(t-1)}\|_2^2 + \frac{\mu_1}{2}\|\vec{\Psi}\vec{x} - \vec{w}_1^{(t-1)} + \vec{u}_1^{(t-1)}\|_2^2$} 
    \label{lin:update-x}
	\STATE{$\vec{v}^{(t)} = \argmin_{\vec{v}}\frac{1}{\sigma^2}\|\vec{A}\vec{x}^{(t)} - \vec{y} + \vec{v}\|_2^2 + \frac{\mu_2}{2}\|g(\vec{v}) - \vec{w}_2^{(t-1)} + \vec{u}_2^{(t-1)}\|_2^2$} 
    \label{lin:update-v}
	\STATE{$\vec{w}_1^{(t)} = \argmin_{\vec{w}_1}  \frac{\mu_1}{2}\|\vec{\Psi}\vec{x}^{(t)} - \vec{w}_1 + \vec{u}_1^{(t-1)}\|_2^2 + \lambda_1\|\vec{w}_1\|_1$} 
    \label{lin:update-w1}
    \STATE{$\vec{w}_2^{(t)} = \argmin_{\vec{w}_2}  \frac{\mu_2}{2}\|g(\vec{v}^{(t)}) - \vec{w}_2 + \vec{u}_2^{(t-1)}\|_2^2 + \lambda_2\|\vec{w}_2\|_1$} 
    \label{lin:update-w2}
    \STATE{$\vec{u}_1^{(t)} =  \vec{u}_1^{(t-1)} + \big(\vec{\Psi}\vec{x}^{(t)} - \vec{w}_1^{(t)}\big)$} 
    \label{lin:update-u1}
    \STATE{$\vec{u}_2^{(t)} =  \vec{u}_2^{(t-1)} + \big(g(\vec{v}^{(t)}) - \vec{w}_2^{(t)}\big)$} 
    \label{lin:update-u2}
	\ENDFOR
	\RETURN{$\widehat{\vec{x}}_{\text{CO}}\gets\vec{x}^{(T)}$} 
  \end{algorithmic}
\end{algorithm}
The subproblem in \linref{update-x} of \algref{core} consists of two quadratic terms and thus has a closed-form solution. For larger problems where explicit matrix representations of $\vec{A}$ and $\vec{\Psi}$ are not feasible, a practical alternative is to solve this problem using gradient descent iterations, with the gradient direction specified by $\frac{2}{\sigma^2}\vec{A}\herm(\vec{A}\vec{x} - \vec{y} + \vec{v}^{(t-1)}) + \mu_1\vec{\Psi}\herm(\vec{\Psi}\vec{x} - \vec{w}_1^{(t-1)} + \vec{u}_1^{(t-1)})$. The subproblem in \linref{update-w1} of \algref{core} admits a closed-form solution given by $\vec{w}_1^{(t)}=\mathcal{S}_{\lambda_1/\mu_1}(\vec{\Psi}\vec{x}^{(t)}+\vec{u}_1^{(t-1)})$, where $\mathcal{S}_{\lambda_1/\mu_1}(\vec{z}) = \frac{\vec{z}}{|\vec{z}|}\max(|\vec{z}| - \lambda_1/\mu_1, \vec{0})$ is the element-wise soft-thresholding, with $|\vec{z}|$ representing the element-wise absolute values. Likewise, the subproblem in \linref{update-w2} of \algref{core} also admits a closed-form solution given by $\vec{w}_2^{(t)}=\mathcal{S}_{\lambda_2/\mu_2}(g(\vec{v}^{(t)}) + \vec{u}_2^{(t-1)})$. 

The gradient direction for the second term in \linref{update-v} of \algref{core} may not be obvious. Here, we outline the gradient direction for this subproblem. By $\vec{r}_2^{(t-1)} \triangleq \vec{w}_2^{(t-1)} - \vec{u}_2^{(t-1)}$, the second term of this subproblem can be presented by
\begin{align}
    H &\triangleq \frac{\mu_2}{2}\|g(\vec{v}) - \vec{r}_2^{(t-1)}\|_2^2 \nonumber \\
    &= \frac{\mu_2}{2}\sum_{j=1}^J \big(\Tilde{v}_j - r_{2,j}^{(t-1)}\big)^{\!*} \big(\Tilde{v}_j - r_{2,j}^{(t-1)}\big)\nonumber \\
    &= \frac{\mu_2}{2}\sum_{j=1}^J \left(\|\vec{v}_j\|_2 - r_{2,j}^{(t-1)}\right)^{\!\!\!*} \left(\|\vec{v}_j\|_2 - r_{2,j}^{(t-1)}\right),
\end{align}
where $r_{2,j}^{(t-1)}$ is the $j^{\text{th}}$ element of $\vec{r}_2^{(t-1)}$ and $(\cdot)^{*}$ represents complex conjugate. The gradient of $H$ using Wirtinger derivatives with respect to $\vec{v}^*_l$,\cite{kreutz2009complex} where $\vec{v}^*_l$ is element-wise complex conjugate of $\vec{v}_l$, is given by
\begin{align}
    \nabla_{l} H = 2\frac{\partial H}{\partial \vec{v}^*_l} &=  2\frac{\partial }{\partial\vec{v}^*_l}\left( \frac{\mu_2}{2}\sum_{j=1}^J \left(\|\vec{v}_j\|_2 - r_{2,j}^{(t-1)}\right)^{\!\!\!*} \left(\|\vec{v}_j\|_2 - r_{2,j}^{(t-1)}\right) \right) \nonumber \\
    &=  \mu_2 \frac{\partial }{\partial\vec{v}^*_l}\left( \left(\|\vec{v}_l\|_2 - r_{2,l}^{(t-1)}\right)^{\!\!\!*} \left(\|\vec{v}_l\|_2 - r_{2,l}^{(t-1)}\right) \right) \nonumber \\
    &=  \mu_2 \left(\vec{v}_l - \frac{1}{2}\left(r_{2,l}^{*(t-1)}+r_{2,l}^{(t-1)}\right)\frac{\vec{v}_l}{\|\vec{v}_l\|_2}\right) \nonumber \\
    &=  \mu_2 \left(\vec{v}_l - \real\big\{r_{2,l}^{(t-1)}\big\}\frac{\vec{v}_l}{\|\vec{v}_l\|_2}\right),
\end{align}
with the gradient across all readouts given by $\nabla H = \left[(\nabla_1 H)\tran, (\nabla_2 H)\tran, \dots,(\nabla_{\!J} H)\tran\right]\tran\in \Complex^{JK\times 1}$. By adding $\nabla H$ to the gradient from the first term in \linref{update-v} of \algref{core}, we arrive at the gradient direction for this subproblem, which is $\frac{2}{\sigma^2}(\vec{A}\vec{x}^{(t)} - \vec{y} + \vec{v}) + \nabla H$.


\clearpage
\section*{Figure S2}
\begin{figure*}[h!]
    \begin{center}
    \includegraphics[width = 1\textwidth]{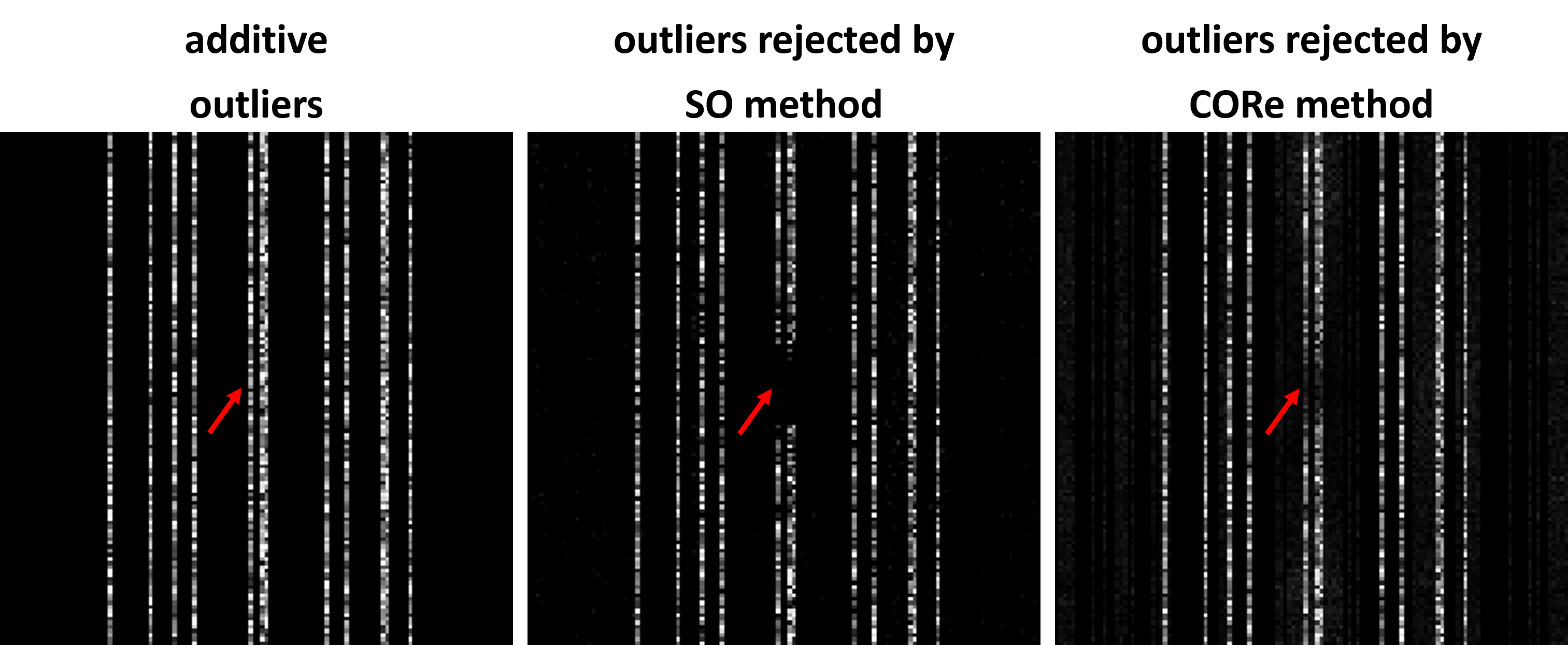}
    \end{center}
    \captionsetup{labelformat=empty}
    \caption{\textbf{FIGURE S2} Comparison of outliers identified by SO (Equation (5)) and \core (Equation (6)) for a representative case in Study I. The left image shows the additive outliers, the middle and the right image display the outliers rejected by SO and \core, respectively. The red arrows highlight that SO is less effective in suppressing the entire outlier readout.}
    \label{fig:Figure S2}
\end{figure*}

\clearpage
\section*{Figure S3}
\begin{figure*}[h!]
    \begin{center}
    \includegraphics[width = 1\textwidth]{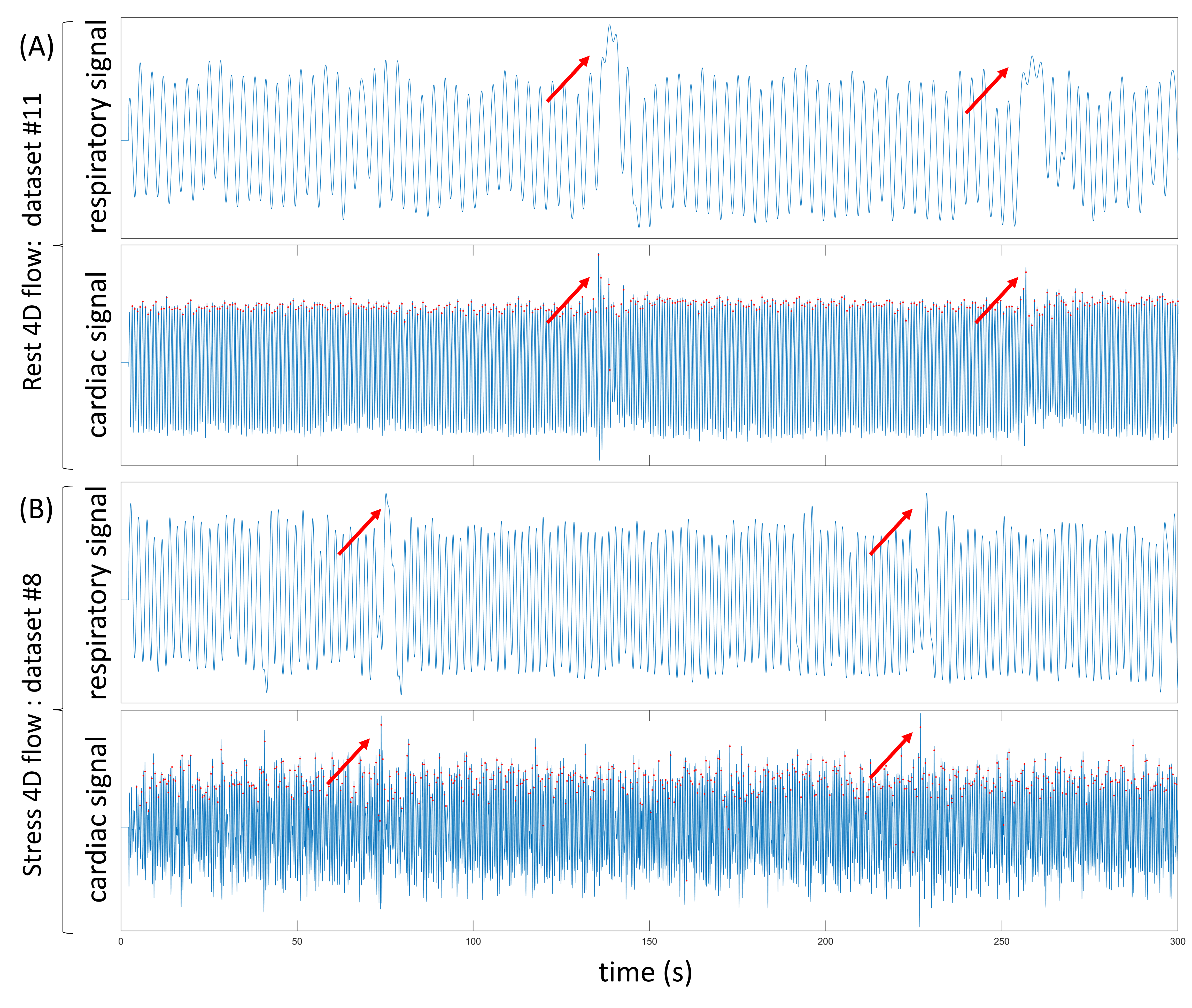}
    \end{center}
    \captionsetup{labelformat=empty}
    \caption{\textbf{FIGURE S3} Cardiac and respiratory motion signals obtained using self-gating from two datasets. (A) displays cardiac and respiratory signals from the rest 4D flow dataset \#11. (B) presents cardiac and respiratory signals from the stress 4D flow dataset \#8. The red arrows highlight specific time points where both the cardiac and respiratory signals appear distorted, potentially indicating subject movement during the scan.}
    \label{fig:Figure S3}
\end{figure*}

\end{document}